\documentclass[12pt,preprint]{aastex}

\shorttitle{Outer halo globular clusters}
\shortauthors{Dotter, Sarajedini, \& Anderson}

\slugcomment{Accepted for publication in The Astrophyical Journal}

\newcommand{\rgc}{\mathrm{R_{GC}}}
\newcommand{\dvi}{\Delta \mathrm{(V-I)}}

\begin{document}

\title{Globular clusters in the outer Galactic halo: new HST/ACS imaging of 6 globular clusters and the Galactic globular cluster age-metallicity relation
\thanks{Based on observations with the NASA/ESA {\it Hubble Space Telescope},
obtained at the Space Telescope Science Institute, which is operated
by AURA, Inc., under NASA contract NAS 5-26555, under program
GO-11586.}}

\author{Aaron Dotter}
\affil{Space Telescope Science Institute, 3700 San Martin Dr., Baltimore, MD 21218}

\author{Ata Sarajedini} 
\affil{Department of Astronomy, University of Florida, 211 Bryant Space Science Center, Gainesville, FL 32611}

\author{Jay Anderson}
\affil{Space Telescope Science Institute, 3700 San Martin Dr., Baltimore, MD 21218}

\begin{abstract}
Color-magnitude diagrams (CMDs) derived from Hubble Space Telescope (HST) Advanced Camera for Surveys $F606W,F814W$
photometry of 6 globular clusters (GCs) are presented. The six GCs form two loose groupings in Galactocentric 
distance ($\rgc$): IC~4499, NGC~6426, and Ruprecht~106 at $\sim$15-20 kpc and NGC~7006, Palomar~15, and Pyxis 
at $\sim40$ kpc. The CMDs allow the ages to be estimated from the main sequence turnoff in every case.
In addition, the age of Palomar~5 ($\rgc \sim 18$ kpc) is estimated using 
archival HST Wide Field Planetary Camera 2 $V,I$ photometry. The age analysis reveals the following: IC~4499, 
Ruprecht~106, and Pyxis are 1-2 Gyr younger than inner halo GCs with similar metallicities; NGC~7006 and Palomar~5 
are marginally younger than their inner halo counterparts; NGC~6426 and Palomar~15, the two most metal-poor GCs in 
the sample, are coeval with all the other metal-poor GCs within the uncertainties. Combined with our previous efforts, 
the current sample provides strong evidence that the Galactic GC age-metallicity relation consists of two 
distinct branches. One suggests a rapid chemical enrichment in the inner Galaxy while the other suggests prolonged GC 
formation in the outer halo. The latter is consistent with the outer halo GCs forming in dwarf galaxies and later being 
accreted by the Milky Way.
\end{abstract}


\section{Introduction}

The Galactic globular clusters (GCs) have long been recognized as useful probes of the formation 
and evolution of the Galaxy. Their spatial coherence allows their locations in age, metallicity, 
and Galactocentric distance (R$_{\rm GC}$) space to be established with some accuracy.
Their numbers allow them to be used as probes of the formation timescale and 
chemical enrichment history of the Milky Way halo, thick disk, and central bulge.

One of the earliest papers to examine the age-metallicity relation (AMR) of the 
Milky Way GC system was \citet{sa89}. They calculated ages for 32 GCs using the 
magnitude difference between the horizontal branch (HB) and the main sequence 
turnoff (MSTO). Taking a cue from \citet{gr85}, \citet{sa89} compared the AMR 
for GCs inside 15 kpc of the Galactic center with that of GCs outside 15 kpc. 
They found a statistically significant difference between the two: the AMR of GCs 
with Galactocentric distance ($\rgc$) $<$ 15 kpc was shallower than that of GCs with 
$\rgc >$ 15 kpc.
\citet{sa89} argued, based on the work of \citet{la72} and \citet{ti78}, that the inner 
halo GCs formed in a region of higher gas and dust density, 
thereby accelerating GC formation and chemical enrichment. In contrast, the outer 
halo GCs formed in lower density sub-galactic fragments in a slower, more chaotic 
fashion akin to the fragmentation and accretion scenario advocated by 
\citet[][see also Navarro et al.\ 1997]{sz78} and featured in models of galaxy 
formation based on the cold dark matter paradigm with a cosmological constant 
\citep[e.g.,][]{ze03, ro05, fo06}.

Subsequent work by \citet{ch92}, \citet{sll}, and \citet{ch96} reinforced 
the results of \citet{gr85} and \citet{sa89} using larger and more reliable 
data sets of GC ages as well as a variety of techniques for measuring them.
The most recent studies to determine ages from a large sample of GCs, by \citet{mf09} 
and \citet{do10}, rely upon the uniform photometric data set provided by the Hubble 
Space Telescope (HST) Advanced Camera for Surveys (ACS) Galactic GC 
Treasury project \citep{sa07}. This data set provides deep photometry for 65 GCs 
of up to $\sim$7 magnitudes below the MSTO \citep{an08}.
Based on a careful analysis of 
the ages and metallicities of the Galactic GC Treasury project GCs, \citet{mf09} 
showed the clearest evidence thus far that the GC AMR splits into two distinct branches. This 
result was reinforced by \citet{do10}, who added the six most distant (known) Galactic GCs to the 
GC Treasury data set and used the combined sample to study the correlations between HB morphology
and a variety of GC parameters.
The analysis presented by \citet{fo10} using data assembled from the literature for 93 
GCs largely reinforces these conclusions.

The HST/ACS Galactic GC Treasury project preferentially targeted nearby GCs in order 
to maximize the depth that could be achieved with one orbit per filter. One consequence of this
strategy is that it left out much of the outer halo.  The most distant GC observed as 
part of the Treasury project is NGC~4147 at $\rgc \sim 21$ kpc. Yet there are many GCs beyond 
this distance that can provide more leverage on the formation and subsequent evolution of 
the outer Galactic halo.

A number of investigators have used HST Wide Field Planetary Camera 2 (WFPC2) photometry to derive 
the ages of GCs with $\rgc > 50$ kpc. \citet{ha97} presented deep photometry of NGC~2419 
\citep[$\rgc \sim 90$ kpc;][2010 revision]{ha96} and concluded that it is coeval with the similarly 
metal-poor GC M\,92 based on a differential comparison of their color-magnitude diagrams (CMDs).
\citet{st99} obtained deep photometry of Palomar~3, Palomar~4, 
and Eridanus (all with $\rgc > 90$ kpc) and concluded that the three are 
$\sim$1.5-2 Gyr younger than the relatively nearby GCs M\,3 and M\,5. Their conclusion rests on the 
assumption that the outer halo GCs have similar chemical abundances ([Fe/H] and [$\alpha$/Fe]) to
M\,3 and M\,5; this assumption has since been borne out for Palomar~3 \citep{ko09} and Palomar~4 \citep{ko10}.
\citet{do08} presented CMDs of Palomar~14 ($\rgc \sim 75$ kpc) and AM-1 ($\rgc \sim 125$ kpc) and concluded 
that both are 1.5-2 Gyr younger than M\,3, in keeping with the results of \citet{st99}.
For the purposes of the present study, the 6 known GCs with R$_{\rm GC} > 50$ kpc
are sufficiently well studied to firmly mark their place in the GC AMR. Chemical abundance information 
remains rather sparse for the most distant GCs, but what is known is consistent with what has been inferred from 
the CMDs.

Thus far then, deep and fairly homogeneous HST CMDs in the $V$ and $I$ (or equivalent) filters exist for the 
majority of GCs 
with $\rgc \la 15$ kpc and all those (known) beyond 50 kpc, allowing precise and internally consistent 
ages to be measured. What remains is to target the GCs within $15 \la {\rm R_{GC}} \la 50$ 
kpc so that the complete radial extent of the Galactic GCs, and what it can tell us about the formation
of the Galaxy, can be probed. The present study is the first major step toward fulfilling this goal, 
wherein we present deep, homogeneous HST ACS photometry, and use the resulting CMDs to estimate the ages, 
of six outer halo GCs: Pyxis, Ruprecht~106, IC~4499, NGC~6426, NGC~7006, and Palomar~15. Of the six, 
only IC~4499 has previous HST photometry \citep{pi02}, and it is not of sufficient depth to allow an 
accurate age determination. In addition, a new age determination for Palomar~5 ($\rgc \sim 18$ kpc) using 
archival HST/WFPC2 $V,I$ data from \citet{gr01} has been performed and included in the subsequent discussion.

The remainder of this paper proceeds as follows: the observations and data reduction are described in 
$\S2$, the CMDs are presented in $\S3$, the available information on the chemical abundances, distances, and 
reddening values of the sample is reviewed in $\S4$, new age determinations based on isochrone fitting are 
presented in $\S5$, the Galactic GC AMR is updated and compared with theoretical work in the context of the 
formation of the Galaxy in $\S6$, $\S7$ provides some discussion of the results, and $\S8$ summarizes the 
findings presented in the paper.

\section{Observations and Data Reduction}

The observations were obtained with the  HST/ACS Wide Field Channel (WFC) under program number 
GO-11586 (PI: Dotter) during Cycle 17. Table \ref{obslog} shows the observing log. The ACS images were 
retrieved from the HST archive and calibrated using the pipeline bias and flat-field procedures. 
All clusters, except for Palomar~15, were imaged over two orbits with the first orbit devoted 
to $F606W$ and the second to $F814W$. Palomar~15 is the most distant and reddened in the sample and 
it received a total of five orbits: two in $F606W$ and three in $F814W$.

\begin{deluxetable}{llllllll}
\tablecolumns{8}
\tablewidth{0pc}
\tabletypesize{\scriptsize}
\tablecaption{GO-11586 Observations \label{obslog}}
\tablehead{\colhead{Cluster}&\colhead{Data set}&\colhead{Date}&\colhead{RA}
          &\colhead{Dec}&\colhead{PA\_V3}&\colhead{F606W}&\colhead{F814W}}
\startdata
Pyxis         & jb1601 & 11/10/2009 & 09h07m57.7s & -37:13:17.0 &  76.86 & 50s, $4\times517$s & 55s, $4\times557$s \\
Ruprecht~106  & jb1602 & 04/07/2010 & 12h38m40.1s & -51:09:00.9 & 287.34 & 55s, $4\times550$s & 60s, $4\times585$s \\
IC~4499       & jb1603 & 01/07/2010 & 15h00m18.5s & -82:12:49.5 & 235.97 & 60s, $4\times603$s & 65s, $4\times636$s \\
NGC~6426      & jb1604 & 04/08/2009 & 17h44m54.6s & +03:10:13.0 & 294.67 & 45s, $4\times500$s & 50s, $4\times540$s \\
NGC 7006      & jb1605 & 05/10/2009 & 21h01m29.4s & +16:11:14.4 & 275.03 & 45s, $4\times505$s & 50s, $4\times545$s \\
Palomar~15    & jb1606 & 16/10/2009 & 16h59m50.9s & -00:32:17.9 & 261.25 & 10s, $4\times550$s & 10s, $4\times500$s \\
\nodata   &        &          &            &             &         & 65s, $4\times550$s & 25s, $4\times560$s \\
\nodata   &        &          &            &             &         &                    & 55s, $4\times525$s 
\enddata
\end{deluxetable}

Photometry and astrometry were extracted from the images using the programs developed
for GO-10775, the ACS Survey of Galactic Globular Clusters, as described by \citet{an08}. 
The observing strategy in the current program was designed to follow the identical pattern 
of exposure times and dithers as GO-10775. Since \citet{an08} give an extremely detailed description 
of the data reduction process, only those aspects that have changed in the interim will be
mentioned here.

Before performing photometry on the individual images, they were corrected for charge transfer efficiency 
using the algorithm developed by \citet{an10}. The instrumental photometry was then calibrated to the HST 
VEGAmag system following \citet{be05} and using the aperture corrections from \citet{si05} as described by
\citet{sa07} and \citet{an08}. The VEGAmag photometric zeropoints were obtained from \citet{bo07}.
The photometric catalogs and supporting data files from the reduction process of the 6 GCs presented herein 
will be made publicly available through the same archive that will host the ACS GC Treasury database.

\section{Color-Magnitude Diagrams}

\begin{figure}
\plotone{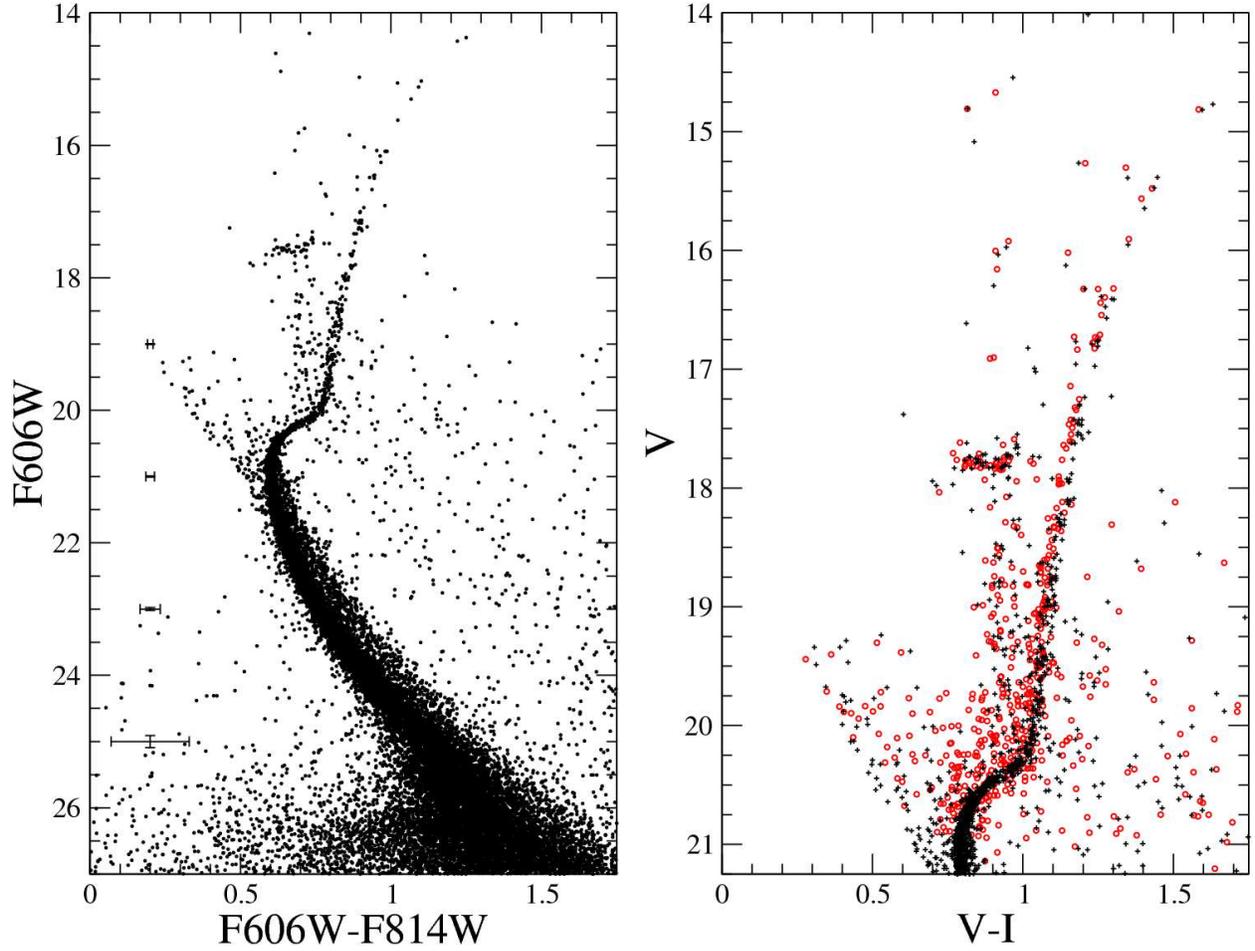}
\caption{Left: The $F606W-F814W$ CMD of Ruprecht~106. Right: The ACS data (crosses), converted to $V$ 
and $I$, compared with the ground-based $V$ and $I$ data of \citet[][open circles]{sl97}.\label{rupr106}}
\end{figure}

\begin{figure}
\plotone{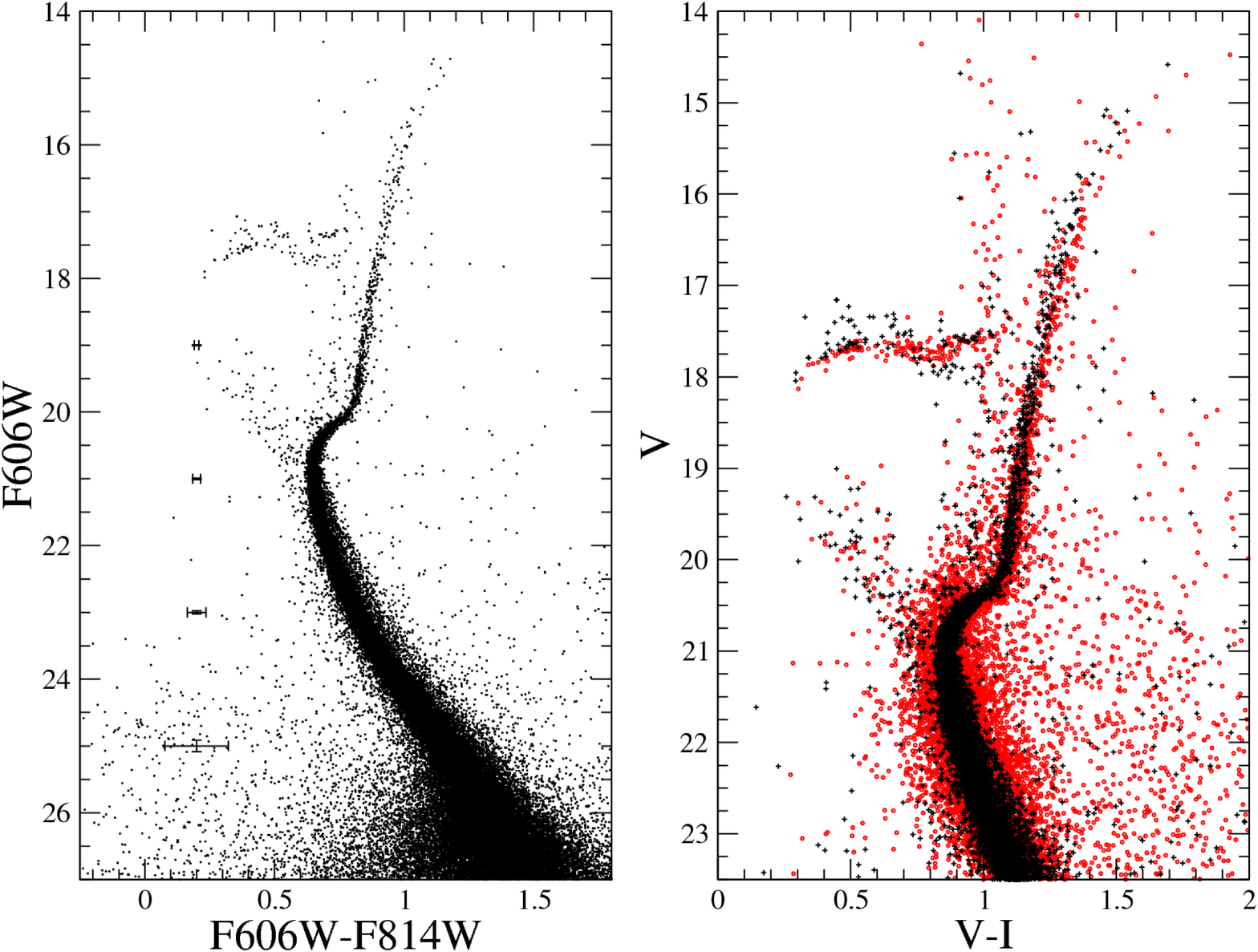}
\caption{Left: The $F606W-F814W$ CMD of IC~4499. Right: The ACS data (crosses), converted to $V$ and $I$, compared with the $V,I$ data of \citet[][open circles]{wa11}.
\label{ic4499}}
\end{figure}

\begin{figure}
\plotone{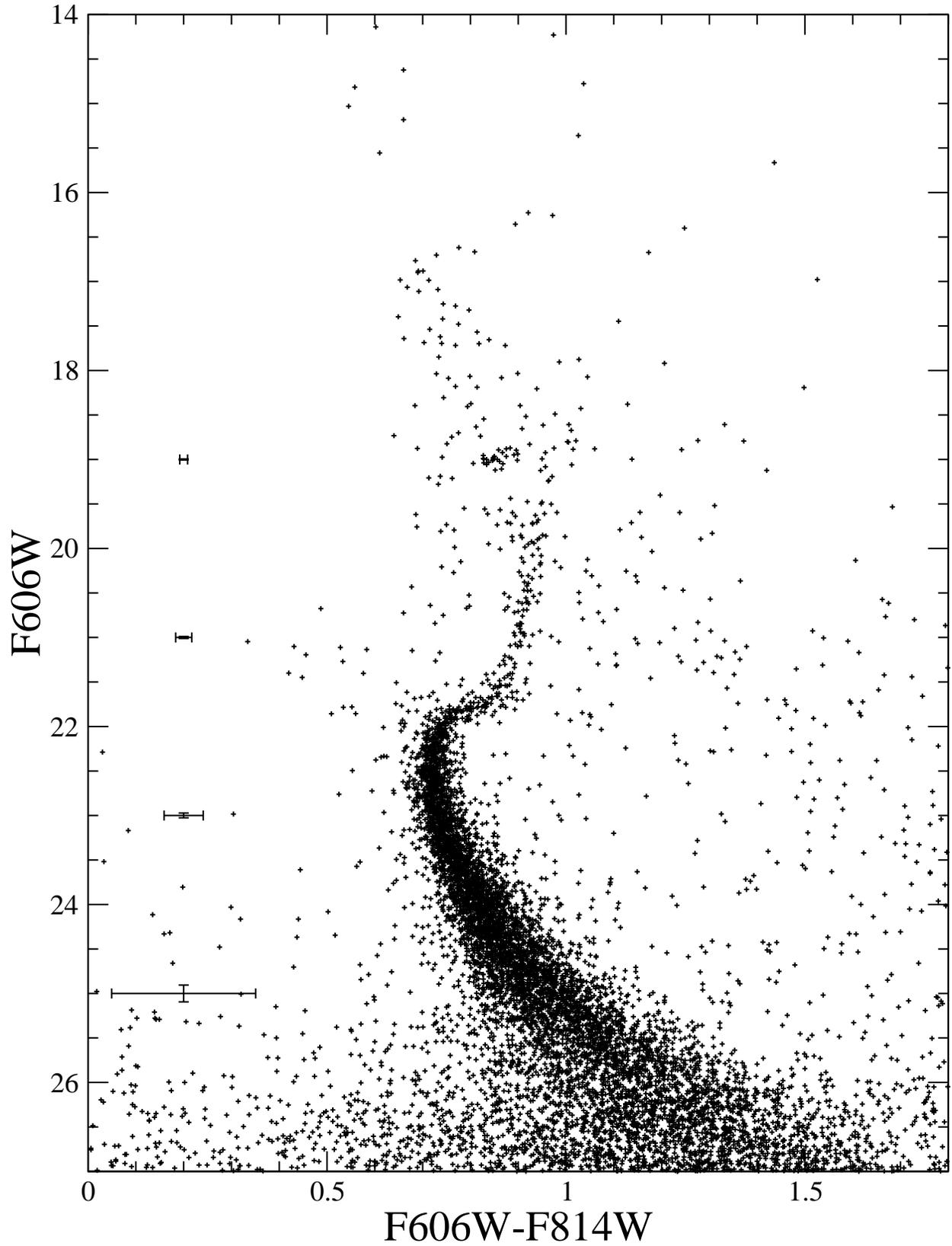}
\caption{The $F606W-F814W$ CMD of Pyxis.\label{pyxis}}
\end{figure}

\begin{figure}
\plotone{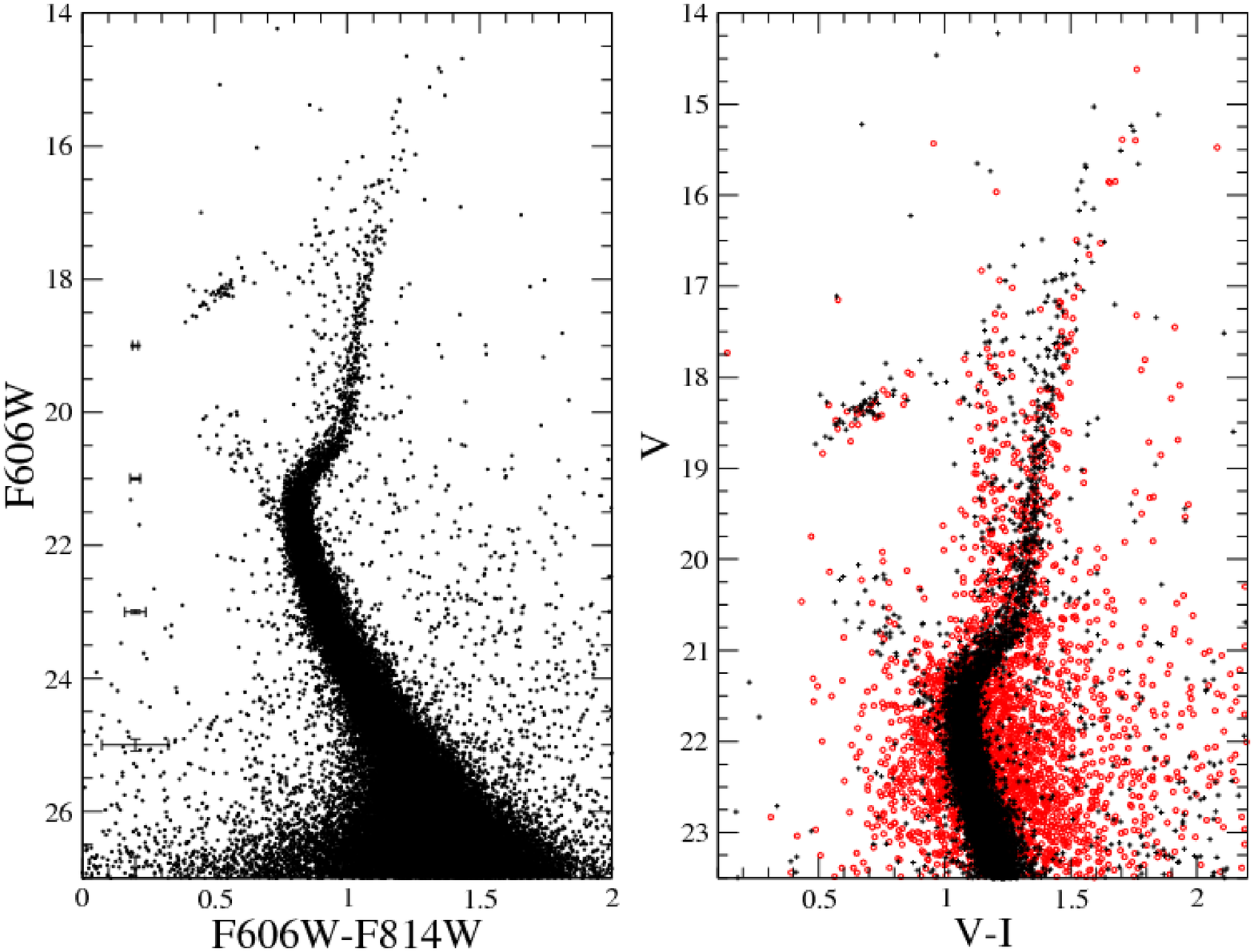}
\caption{Left: The $F606W-F814W$ CMD of NGC~6426. Right: The ACS data (crosses), converted to $V$ and $I$, compared
with the $V,I$ data of \citet[][open circles]{ha99}.\label{ngc6426}}
\end{figure}

\begin{figure}
\plotone{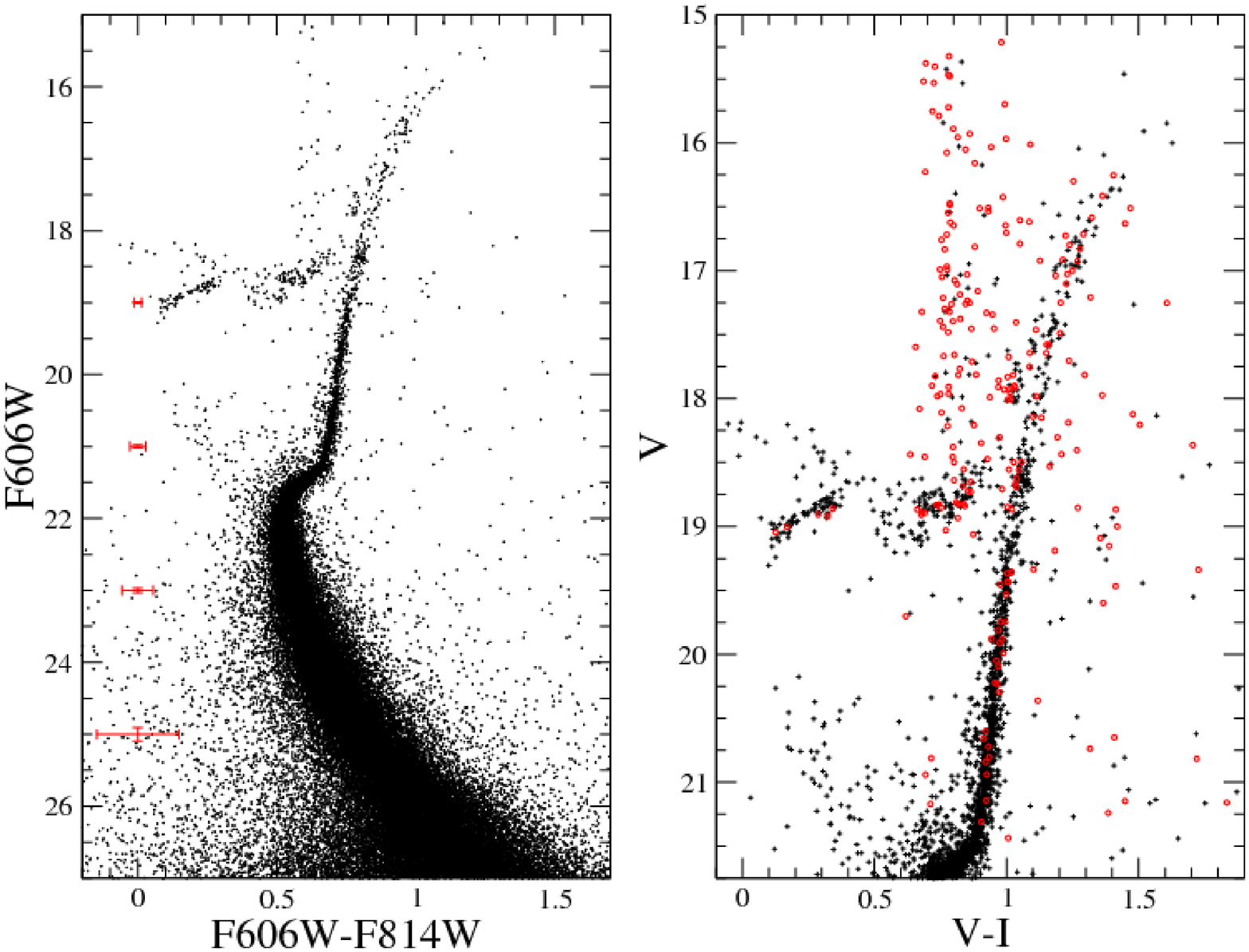}
\caption{Left: The $F606W-F814W$ CMD of NGC~7006. Right: The ACS data (crosses), converted to $V$ and $I$, compared
with the $V,I$ data of \citet[][open circles]{st00}.\label{ngc7006}}
\end{figure}

\begin{figure}
\plotone{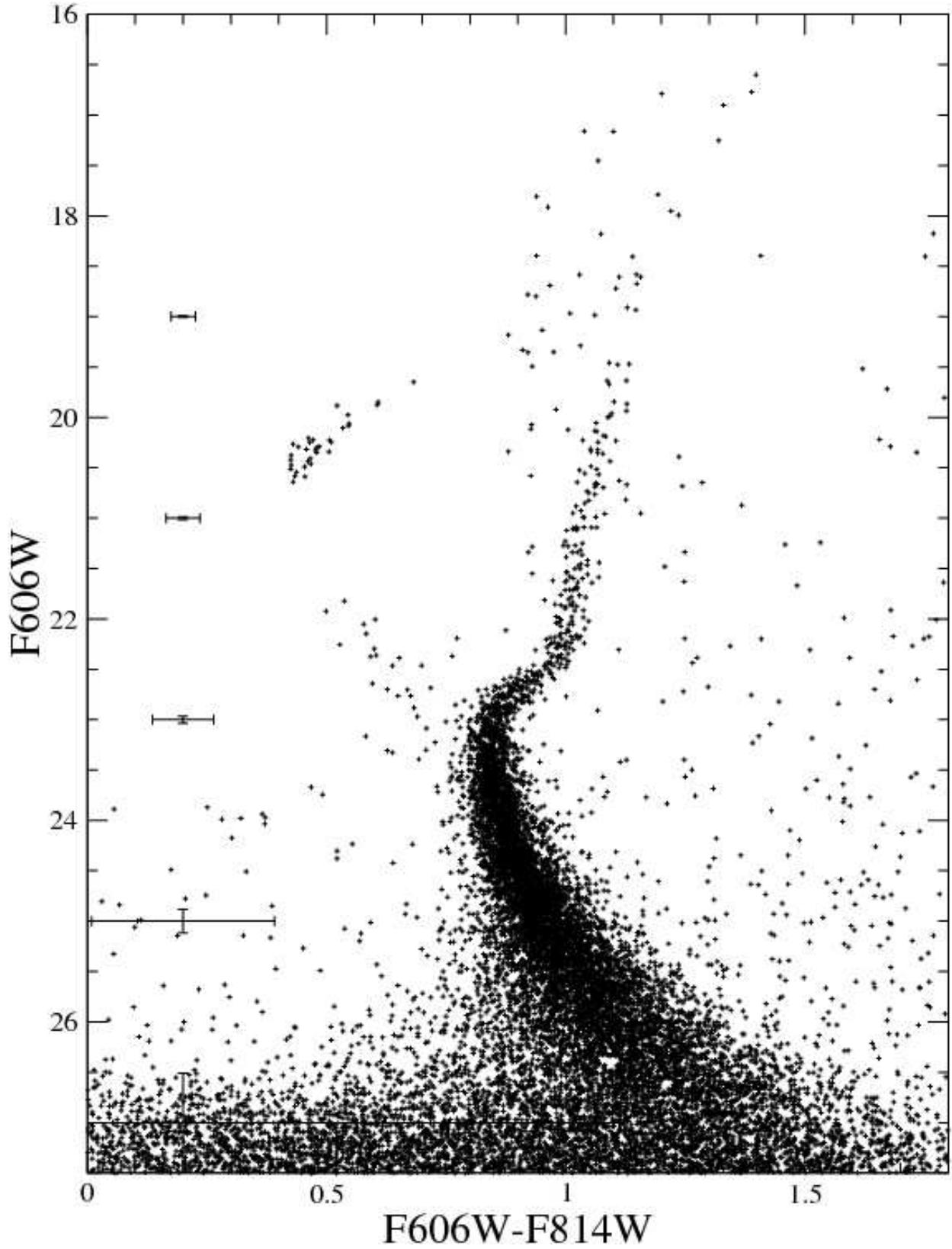}
\caption{The $F606W-F814W$ CMD of Palomar~15.\label{palmr15}}
\end{figure}

The $F606W,F606W-F814W$ CMDs are presented in Figures \ref{rupr106} through \ref{palmr15}.
Typical photometric errors are demonstrated towards the left side of each figure.
In cases where ground-based $V,I$ data are available, we also make a direct comparison
with the ACS data converted to $V$ and $I$ using the empirical transformations provided by 
\citet{si05}.
The ground-based comparisons indicate, in all cases, that the HST photometric system maintains 
a high degree of homogeneity post-SM4 and that the transformations to ground-based $V$ and $I$ 
magnitudes perform well.

The CMDs reveal several striking features. In particular, it can be seen that Ruprecht~106
and IC~4499 have strong binary and blue straggler sequences. 
NGC~6426, Palomar~15, and Pyxis show signs of differential reddening and, indeed, these 
are the most reddened clusters in the present sample. Palomar~15 and Pyxis are sparsely 
populated clusters; the red giant branch (RGB) of Pyxis loses coherence for $F606W \la 19$.

\section{Previous studies}

This section provides a brief review of noteworthy studies targeting the GCs considered in the present study. A few basic parameters
are collected in Table \ref{basic}.

\begin{deluxetable}{llllll}
\tablecolumns{6}
\tablewidth{0pc}
\tablecaption{Literature values of basic cluster parameters\label{basic}}
\tablehead{\colhead{Cluster}&\colhead{DM$_\mathrm{V}$}&\colhead{$E(B-V)$}&\colhead{$E(B-V)$}&\colhead{[Fe/H]}&\colhead{[$\alpha$/Fe]}\\
           \colhead{}       &\colhead{(Harris)}        &\colhead{(Harris)}  &\colhead{(SFD98)}   &\colhead{(Harris)}&\colhead{}}
\startdata
IC~4499        & 17.09  &  0.23 &  0.224 & $-1.6$  &   \nodata    \\
Ruprecht~106   & 17.25  &  0.20 &  0.174 & $-1.5^a$   &  $\sim0.0^a$ \\
NGC~6426       & 17.70  &  0.36 &  0.346 & $-2.26$  &   \nodata    \\
NGC~7006       & 18.24  &  0.05 &  0.082 & $-1.55^b$  &  $+0.23^b$   \\
Pyxis 	       & 18.65  &  0.21 &  0.327 & $-1.4^c$&  \nodata     \\ 
Palomar~15     & 19.49  &  0.40 &  0.394 & $-2.0^d$  &   \nodata    \\
Palomar~5      & 16.92  &  0.03 &  0.056 & $-1.3^e$  &   $+0.16^e$  \\
\enddata
\tablerefs{Sources of data are listed in the column headers except where otherwise specified. 
Harris--\citet{ha96}; SFD98--\citet{sc98}; ($a$) P.\ Francois (2010) priv. comm.; ($b$) \citet{kr98}; ($c$) \citet{pal00}; 
($d$) \citet{da95}; ($e$) \citet{sm02} }
\end{deluxetable}

\subsection{NGC~6426}
The first modern CMD of this cluster was presented by \citet{zi96}
based on imaging with the 1-m telescope at Cerro Tololo
Inter-American Observatory. Their CMD extended $\sim$3 mags below the
HB and confirmed the identity of 12 variable stars in the cluster.
\citet{ha99} and \citet{pa00} published $BVRI$ photometry extending to 
about 1 mag below the MSTO and an extensive study of its RR Lyrae variable 
population, respectively. \citet{ha99} concluded that NGC~6426 
is coeval with the mean age of the metal-poor Galactic GCs.

\subsection{IC~4499}
The first CCD photometry of IC~4499 was published by \citet{sa93} 
wherein a $B-V$ CMD extending $\sim$3 mags below the HB was
presented. IC~4499 is exceptional for its prodigious RR Lyrae
population. \citet{su91} compiled luminosity-normalized RR Lyrae 
populations for 77 Galactic GCs.\footnote{
The quantity used by \citet{su91}, N$_{RR}$, is an integrated luminosity-averaged quantity: 
the number of RR Lyraes a GC would have if the GC itself had $M_V=-7.5$.}
On a scale where N$_{RR}$ = 0.3 for 47~Tuc and N$_{RR}$ = 56.4 for
M\,3, the N$_{RR}$ value of IC~4499 is 88.7. The only GC with a higher
value of N$_{RR}$ is Palomar~13, but with only four RR Lyrae stars 
its specific frequency is likely to be influenced by small number statistics. 

A comprehensive study of the CMD and
RR Lyrae population of IC~4499 was performed by \citet{wn96}, who
derived a mean reddening of E(B$-$V)=$0.22\pm0.02$ based on four
independent estimates and [Fe/H]= $-1.65\pm0.10$ on the \citet{zw84} 
scale. Adopting a reddening and metallicity close to these values, 
\citet{fe95} conclude that IC~4499 is younger (by 3-4 Gyr) than most 
globular clusters at its metallicity.

\citet{ku11} presented an updated study of the RR Lyraes in 
IC~4499 and found that their period change rates were an order of magnitude
larger than predicted by stellar models. \citet{wa11} presented multi-band, 
ground-based photometry of IC~4499. Their CMD reaches $\sim 3$ mag below the 
MSTO from which \citet{wa11} estimated an age of $12\pm1$ Gyr.

\subsection{Ruprecht~106}
Discovered by Ruprecht on photographic plates as he was searching for
open clusters \citep{al61}, Ruprecht~106 existed in relative
obscurity until the first CCD photometry by \citet{bu90} showed
that it is metal-poor with a red HB and, thus, was likely to be younger 
than other GCs with similar metallicities. Subsequent work by \citet{bu93} 
corroborated this result.

The metal abundance of Ruprecht~106 has been a source of controversy 
since the work of \citet{bu90}.
The abundance studies performed thus far have fallen into two broad
categories, those that find a metal abundance of [Fe/H] $\sim -1.9$
\citep{bu90,bu93,sl97} based on photometric indicators and those that 
conclude [Fe/H] $\sim -1.6$ based on spectroscopy \citep{da92,fr97}. 
\citet{ka95} found that Ruprecht~106 and M\,3 have similar metallicities
based on time series photometry of 12 RR Lyrae variables and, separately,
that [Fe/H] $\geq -1.6$ based on the relative locations of the RGB bump
and the HB.

The discrepancy in metallicity estimates may be due to the unusually low
[$\alpha$/Fe] ratio of Ruprecht~106 compared to other metal-poor Galactic 
GCs \citep{pvi05}. \citet{bw97} claimed [Fe/H]=$\simeq -1.45$ and [O/Fe] $\sim 0$ based 
on high-resolution spectra of 2 red giants. VLT/UVES spectroscopy of 6 red 
giants indicates $-1.5 < $ [Fe/H] $ < -1.45$ and $-0.1 < $ [$\alpha$/Fe] 
$< +0.1$ (P.\ Francois, 2010, private communication).

\subsection{Palomar~15}

The earliest CCD photometry of Palomar~15 dates back to the work of
\citet{sc90} and \citet{ha91}. CMDs presented in both 
of these studies extend from the tip of the RGB to about 3 mags below
the HB and reveal an HB morphology that is blue-ward of the RR Lyrae
instability strip. In addition, both studies conclude that Palomar~15 
suffers from a higher-than-expected amount of line-of-sight reddening.
Whereas the \citet{bh82} reddening maps predict E(B$-$V) $\sim$ 0.1, 
the CMDs of \citet{sc90} and \citet{ha91} suggest a reddening closer 
to E(B$-$V) $\sim$ 0.4, consistent with the reddening maps of \citet{sc98}.
Given this value, photometric indicators predict [Fe/H] $\sim$ -1.9. 
The only spectroscopic value of the metallicity was published by \citet{da95}, 
who found [Fe/H]= $-2.00\pm0.08$ based on the strength of its Calcium triplet
lines.

\subsection{NGC~7006}
NGC~7006 was recognized early-on as a prime example of
the second parameter effect. \citet{sa67} noted that
that M\,13, M\,3, and NGC~7006 all have very similar metallicities
([Fe/H] $\sim -1.6$) but exhibit a range of HB morphologies,
with M\,13 displaying the bluest HB and NGC~7006 the reddest.
Despite its importance in this regard, the deepest published CMD 
remains that of \citet{bu91}, which extends only $\sim$1 mag below
the MSTO.
\citet{kr98} estimated [Fe/H]= $-1.55$ based on high-resolution 
spectra of 6 stars. \citet{ki08} found similar a similar average value
from medium-resolution spectra of 20 stars: [Fe/H]= $-1.59\pm0.03$.

\subsection{Pyxis}
Discovered by \citet{we95} while searching the optical sky surveys,
the Pyxis globular cluster was confirmed as a GC by \citet{dc95}
and, soon thereafter, by \citet{ir95} and \citet{sg96}. These studies
revealed a sparsely populated CMD indicative of an intermediate metallicity
GC with a red HB.
\citet{sg96} used the simultaneous reddening and metallicity method of 
\citet{sa94} to derive [Fe/H] = $-1.20\pm0.15$ and E(B$-$V) = $0.21\pm0.03$
for Pyxis. Spectroscopy of 6 bright giants in Pyxis by \citet{pal00}
yields a mean metallicity of [Fe/H] = $-1.4\pm0.1$, broadly
consistent with the abundance derived by \citet{sg96}.

\subsection{Palomar~5}
The first CCD-based photometry of Palomar~5 was published by \citet{sm86} in
the $B$ and $V$ filters. As is typical for the Palomar clusters, the CMD of
Palomar~5 is relatively sparse on the RGB and HB. Despite the fact that their CMD
clearly delineated the MSTO of Palomar~5, \citet{sm86} were unable to determine a reliable
age estimate for the cluster using the isochrones of \citet{vb85}. \citet{gr01} 
presented a deeper, more precise CMD of Palomar~5 based on HST/WFPC2 
observations in the $F555W$ and $F814W$ filters, which they  converted to 
ground-based $V,I$. The resulting CMD extends over 7 magnitudes below the 
MSTO and was used by \citet{gr01} to study the main sequence luminosity function 
of Palomar~5. Since the study of \citet{gr01}, Palomar~5 has been the subject of extensive
investigations focused on its tidal tails \citep{gr06,od09}. Spectroscopy of four giant
stars in Palomar~5 yields a mean metallicity of [Fe/H] $\sim -1.3$ \citep{sm02}
which is consistent with the abundance inferred from the CMD \citep{sm86}.
The reddening of Palomar~5 is thought to be relatively low, as noted by \citet{sm86},
who quote a value of E($B-V$) = 0.03. The maps of \citet{sc98} give a reddening
of E($B-V$) = 0.056 in the direction of Palomar~5.

\section{Age determinations from isochrone fitting}
\citet{do10} presented isochrone fits to the photometric catalog of the ACS Survey of Galactic GCs, 
excluding only those known at the time to possess multiple, distinct stellar populations. The fits
were performed with isochrones from \citet{do08}. The procedure employed in the isochrone fits presented 
in this section follows that of \citet[][Sec. 4.2]{do10}. The distance, reddening, and chemical composition 
of the isochrones are initially set to the best available literature values (see Table \ref{basic}) 
and these are adjusted only when necessary to achieve the best simultaneous agreement between the models and 
the CMD on the unevolved main sequence and the base of the red giant branch. 
\citet[][see their Figures 6 and 7]{do10} showed that the distances derived in this manner were consistent with 
the absolute magnitudes of HB stars derived from RR Lyraes with the $1\sigma$ uncertainties and that the derived 
metallicities were typically within 0.1-0.2 dex of spectroscopically determined values, particularly in the range 
$-2 \leq$ [Fe/H] $\leq -1.5$. The values derived from isochrone fitting for the present sample are compared to the 
values obtained from the literature at the end of this section.

Once the isochrones were registered in this manner, the age is estimated by selecting the
isochrone which most closely matches the shape and position of the subgiant branch. The isochrone fits are performed
`by eye' and the quoted uncertainties reflect the width of the data in the CMD and any mismatch between the 
data and the slope of the models through this region. As such, these uncertainties are suggestive of goodness-of-fit
but lack statistical significance. \citet{ck02} report that uncertainties in stellar models lead to an error of $\pm3\%$
in derived ages if a star's properties are known exactly. At an age of 13 Gyr, 3\% corresponds to $\sim0.4$ Gyr. Adding 
this value in quadrature with a typical uncertainty of 1 Gyr due to the fit (see Table \ref{agemet}) and $\sim0.5$ Gyr for 
0.2 dex uncertainty in metallicity \citep[e.g.,][]{do08} leads to a realistic error budget of $\sim1.2$ Gyr.

As can be seen in the figures that follow, age effects are only relevant from 
$\sim2$ magnitudes below to $\sim1$ magnitude above the MSTO: greater depth in the CMD provides a 
tighter constraint on the models.  The CMDs of Ruprecht~106, IC~4499, and NGC~7006 provide excellent constraints;
differential reddening is a limiting factor in NGC~6426, Palomar~15, and Pyxis.  The sparseness of the Pyxis 
and Palomar~15 CMDs also reduces the precision of the isochrone fitting. 

Stars brighter than the subgiant branch are saturated in the WFPC2 observations of Palomar~5.
To enable a more robust age determination, the WFPC2 data have been supplemented with ground-based observations 
from \citet{st00}. For the $\sim$20 stars that appear in both catalogs, the average offsets are less than 0.01 mag in $V$
and $I$; the WFPC2 data were adjusted to match the photometric zeropoints of the standard stars measured by \citet{st00}.

\begin{figure}
\includegraphics[width=0.3\textwidth]{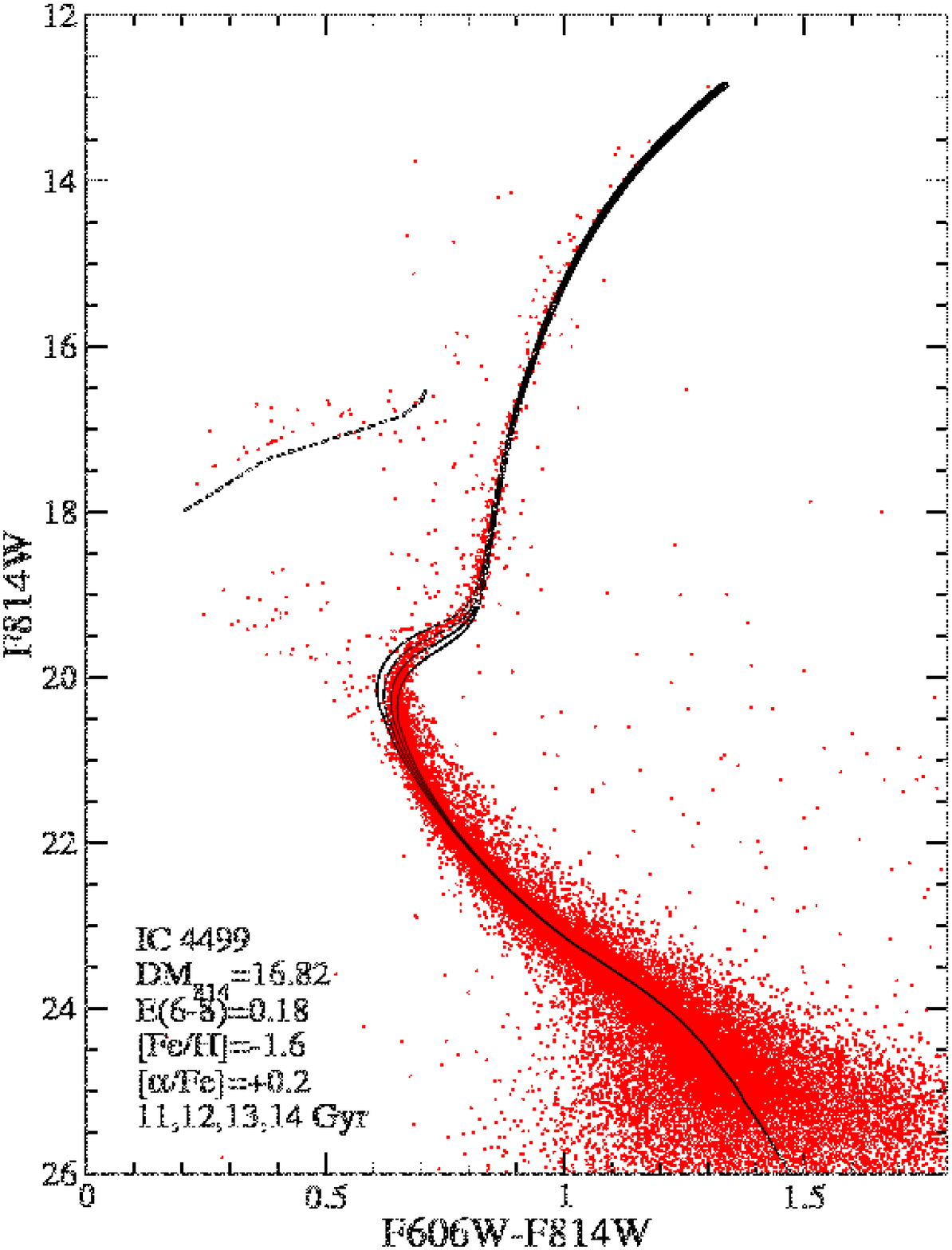}
\includegraphics[width=0.3\textwidth]{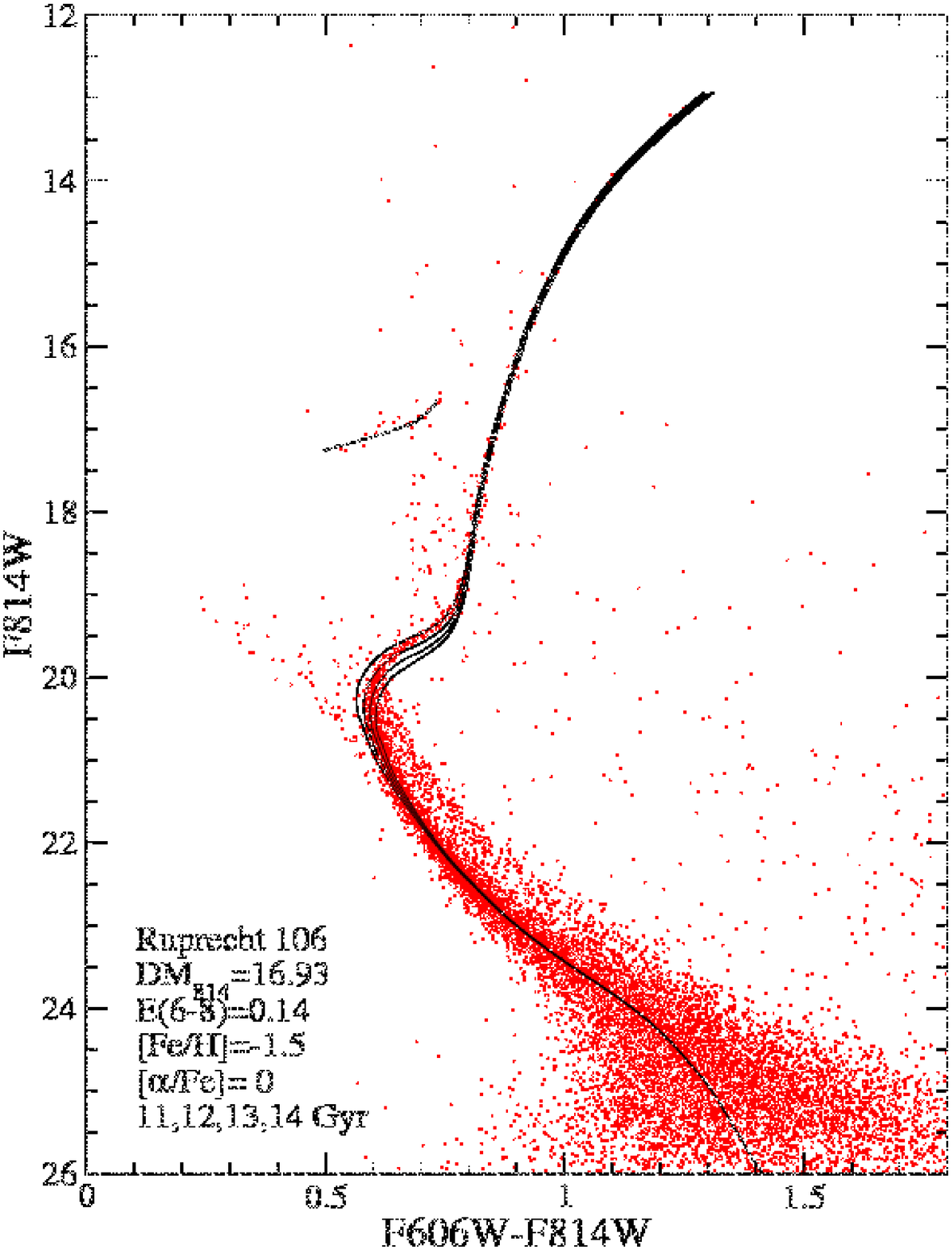}
\includegraphics[width=0.3\textwidth]{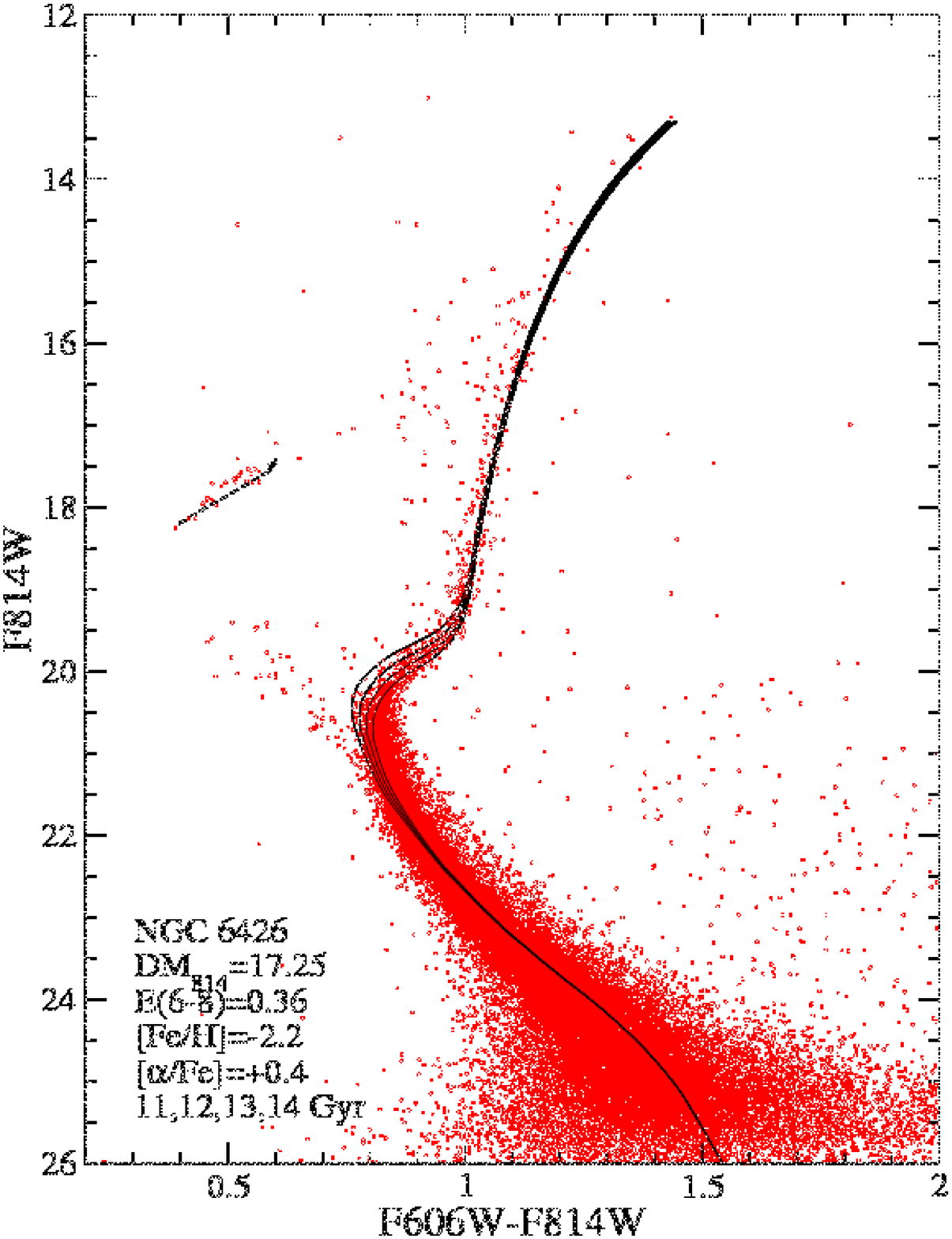}
\caption{GCs at 15-20kpc.\label{iso1}}
\end{figure}

\begin{figure}
\includegraphics[width=0.3\textwidth]{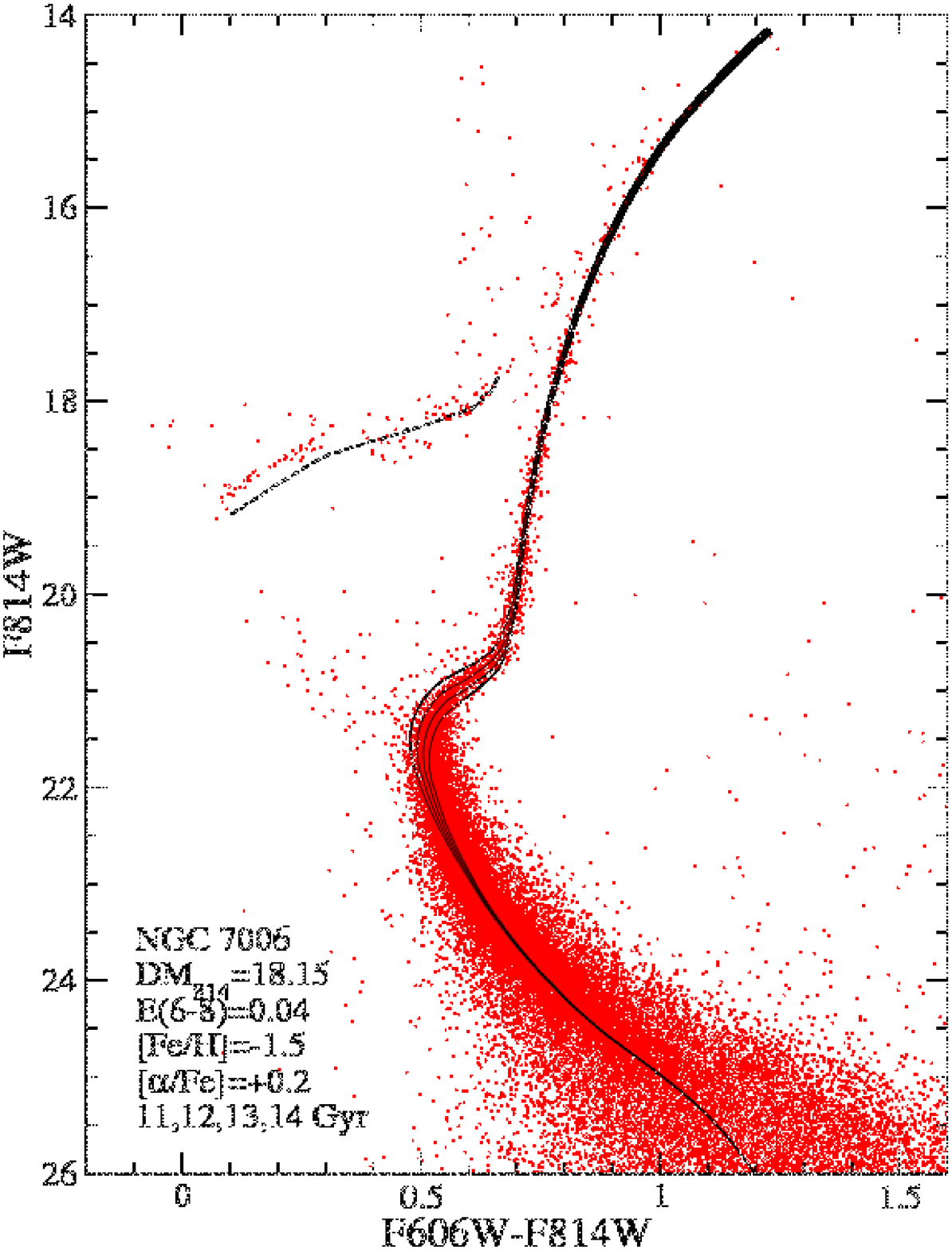}
\includegraphics[width=0.3\textwidth]{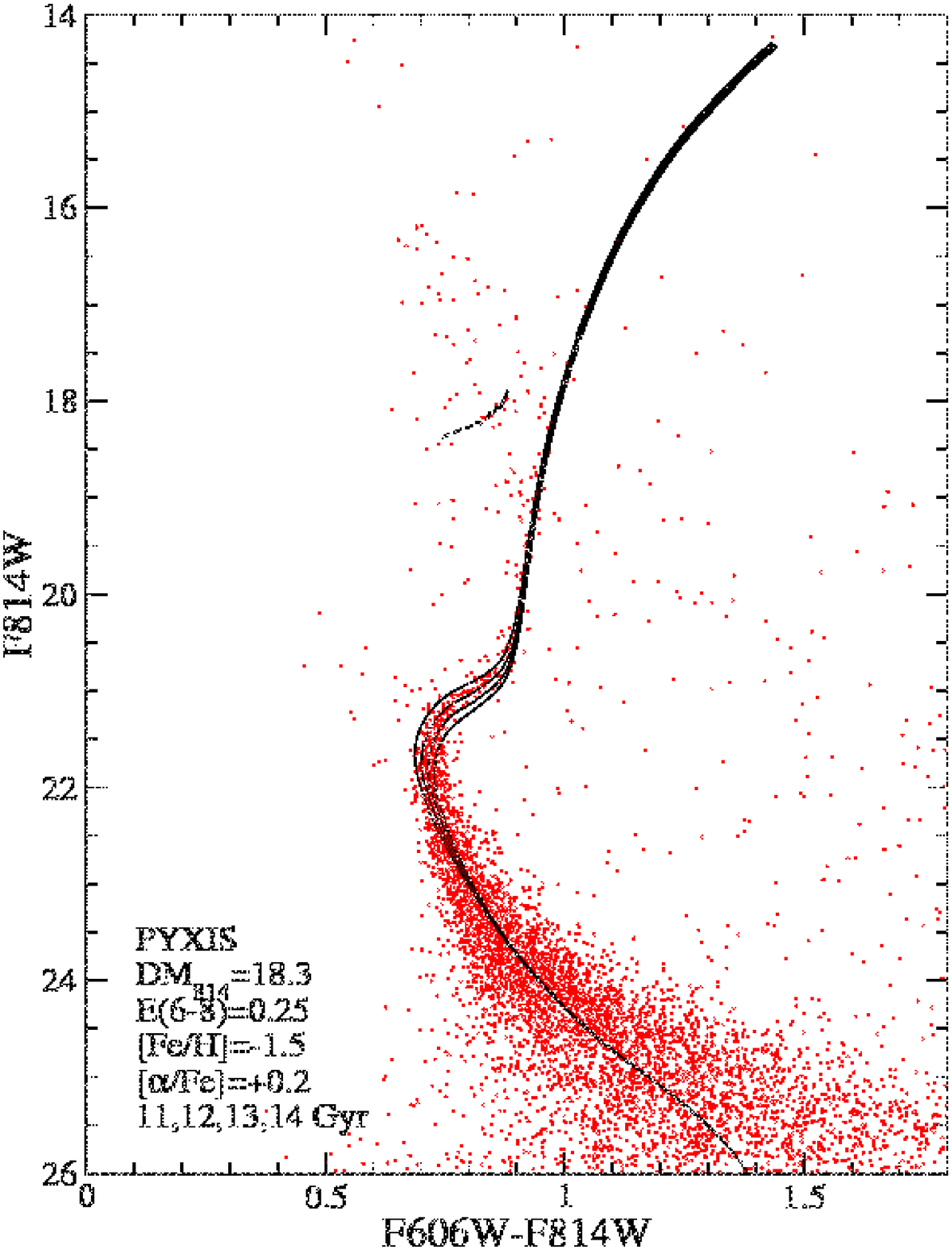}
\includegraphics[width=0.3\textwidth]{isoPal15v2BMP}
\caption{GCs at $\sim40$kpc.\label{iso2}}
\end{figure}

\begin{figure}
\plotone{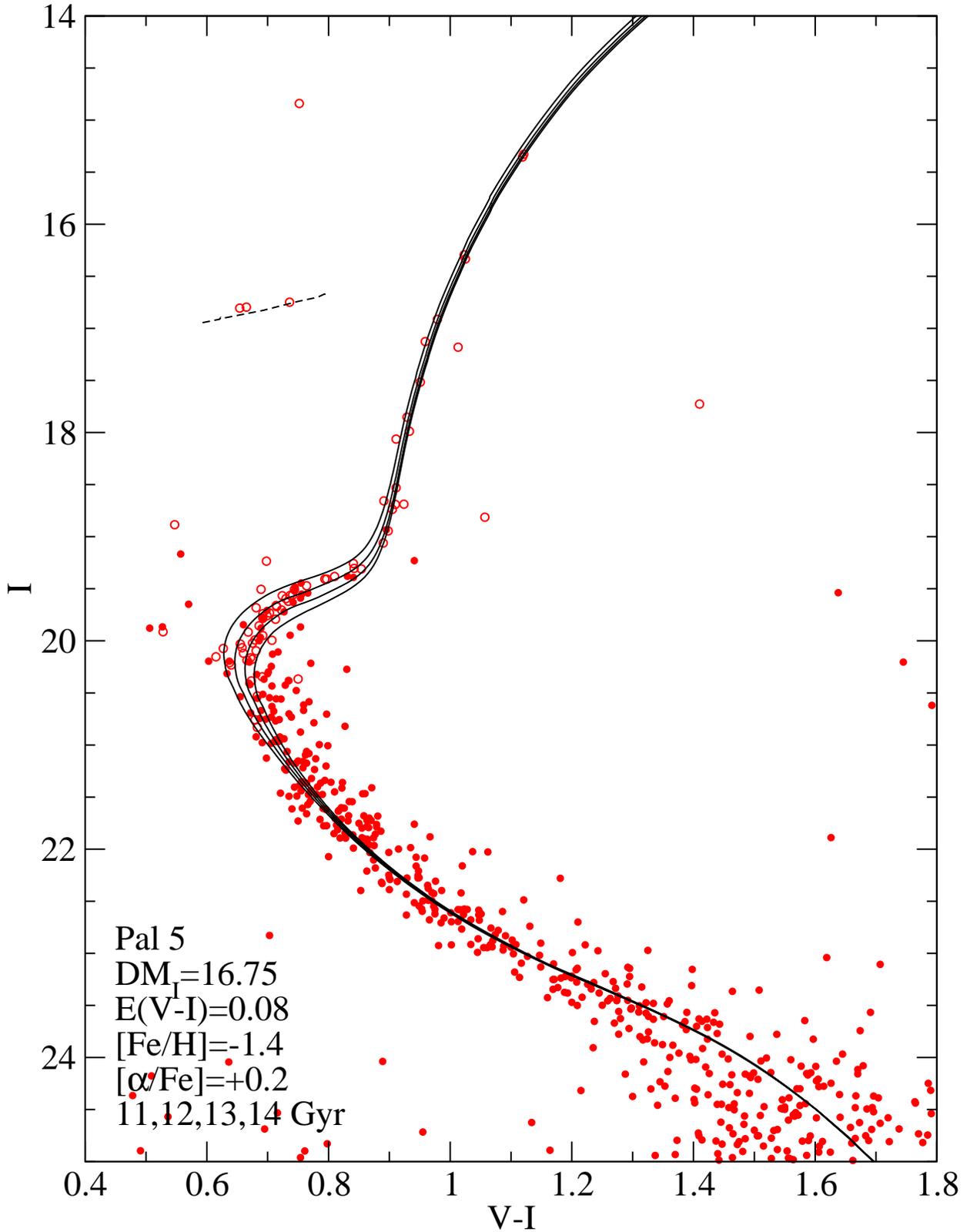}
\caption{Isochrone fits to the HST/WFPC2 CMD of Palomar~5 in the ground $V,I$ system.
The WFPC2 data are shown as filled circles and the ground-based data as open circles.
\label{isoPal5}}
\end{figure}

\begin{deluxetable}{llllll}
\tablecolumns{6}
\tablewidth{0pc}
\tablecaption{Results of isochrone fitting \label{agemet}}
\tablehead{\colhead{Cluster}&\colhead{DM$_\mathrm{V}$}&\colhead{$E(B-V)$}&\colhead{[Fe/H]}
          &\colhead{[$\alpha$/Fe]}&\colhead{Age (Gyr)}}
\startdata
IC~4499      & 17.07  &  0.18 & $-1.6$  &   0.2  &  $12.0\pm0.75$ \\
Ruprecht~106 & 17.12  &  0.14 & $-1.5$  &   0.0  &  $11.5\pm0.5 $ \\
NGC~6426     & 17.74  &  0.36 & $-2.2$  &   0.4  &  $13.0\pm1.5 $ \\
NGC~7006     & 18.16  &  0.04 & $-1.5$  &   0.2  &  $12.25\pm0.75$\\
Pyxis 	     & 18.64  &  0.25 & $-1.5$  &   0.2  &  $11.5\pm1.0 $ \\
Palomar~15   & 19.50  &  0.38 & $-2.0$  &   0.4  &  $13.0\pm1.5 $ \\
Palomar~5    & 16.86  &  0.08 & $-1.4$  &   0.2  &  $12.0\pm1.0 $
\enddata
\end{deluxetable}

The isochrone fits are shown in Figures \ref{iso1}, \ref{iso2}, and \ref{isoPal5}; the results of isochrone
fitting are presented in Table \ref{agemet}.  For comparison purposes, the reddening and distance values 
listed in Table \ref{agemet} have been transformed to standard DM$_\mathrm{V}$ and $E(B-V)$ using the extinction
coefficients given by \citet{si05}. The CMDs in Figures \ref{iso1} and \ref{iso2} were cleaned to 
improve the clarity of the fits by selecting stars according to photometric errors and other diagnostic 
information as described by \citet[][Section 7]{an08}. As check on the results of isochrone fitting, synthetic zero age 
HB model sequences for the color ranges appropriate for each GC are plotted in Figures \ref{iso1}-\ref{isoPal5} 
as dashed lines. Generally speaking, these comparisons lend confidence to the results of the isochrone fitting. However, 
note that the models appear to inadequately represent those GCs that harbor both red and blue HB stars and, further, 
that the scarcity of HB stars in Pyxis and Palomar~5 make these comparisons of relatively little use. 

Comparison of the [Fe/H] values listed in Tables \ref{basic} and \ref{agemet} indicate that the literature values
and those adopted in the isochrone fits differ, at most, by 0.1 dex in [Fe/H]. The largest departures are in cases where
no information regarding [$\alpha$/Fe] is available. Comparison of the derived distance moduli shows that 4 of 7 GCs
considered here differ by less than 0.05 mag. The largest discrepancy in distance modulus, that of Ruprecht~106, also 
corresponds to the largest discrepancy in [Fe/H] between the \citet{ha96} catalog and the adopted spectroscopic value. 
The Harris catalog distances are based a the relation between the absolute magnitude of the RR Lyrae stars and the
mean [Fe/H] of the GC.

\section{The age-metallicity relation of Galactic globular clusters}

As described in $\S2$, all aspects of the observations and data reduction of the 6 GCs 
observed with ACS in program GO-11586 were designed to be homogeneous with the ACS Survey 
of Galactic GCs. The observations and 
data reductions of Palomar~5 are consistent with the outer halo GCs studied by \citet{st99}
and \citet{do08}. The isochrone analysis of all 7 GCs presented herein is consistent with 
the approach used in \citet{do08} and \citet{do10}. Hence, the results in Table \ref{agemet} 
can be incorporated into the AMR assembled by \citet{do10}.

\begin{figure}
\plotone{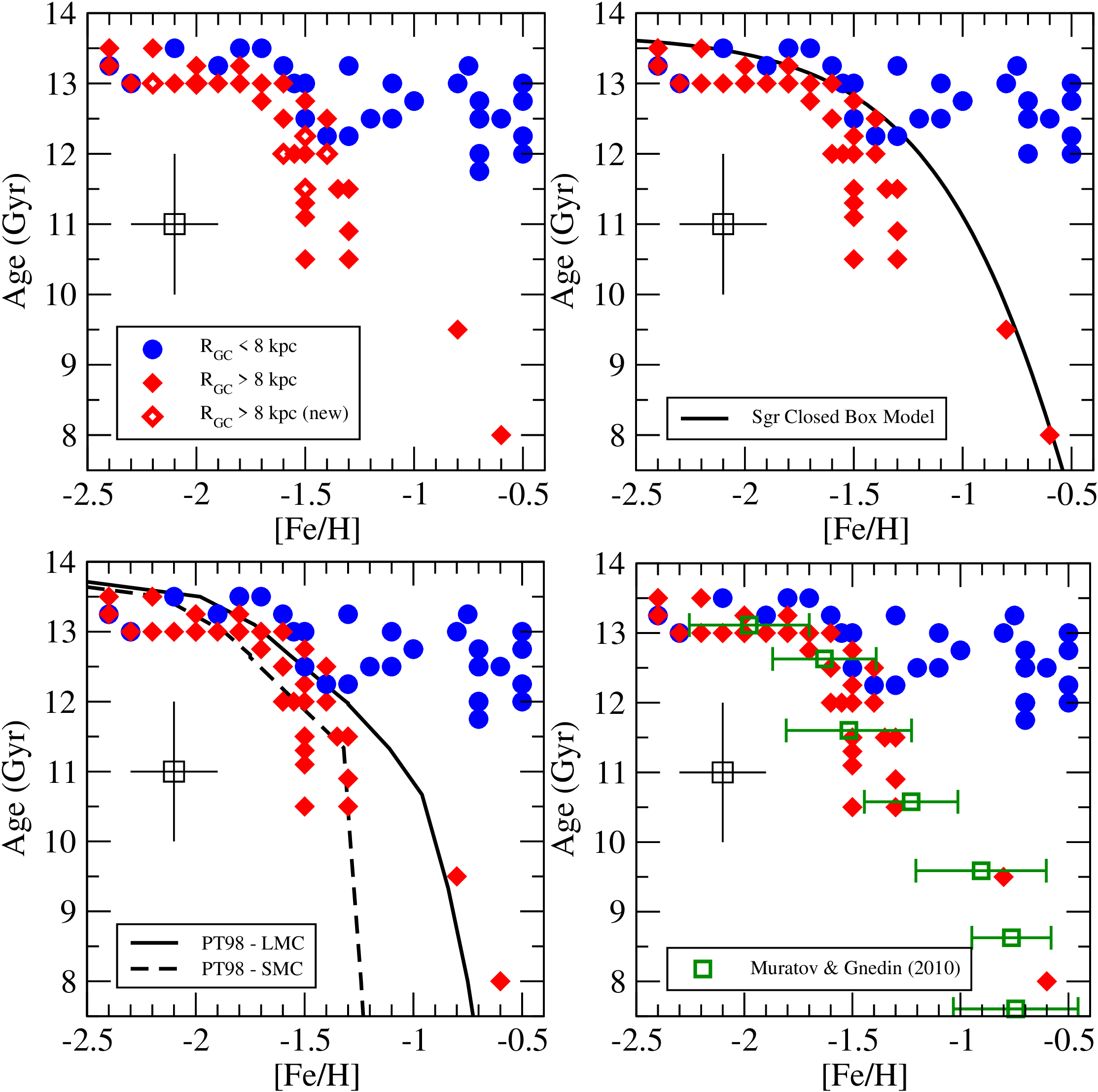}
\caption{{\it Upper left:} The AMR from \citet{do10} with points identifying individual GCs by $\rgc \leq 8$ kpc 
(filled circles) and $\rgc > 8$ kpc (filled diamonds). The current sample are shown as open diamonds.
{\it Upper right:} The AMR with the M\,54/Sagittarius closed box model. 
{\it Lower left:} The AMR with the LMC and SMC models from \citet{pa98}.
 {\it Lower right:} The AMR with the mean trend from the \citet{mg10} models.
The GC symbols are the same as in the upper-left panel except that the current sample is not highlighted. See text 
for discussion.
\label{AMR}}
\end{figure}

The upper-left panel of Figure \ref{AMR} shows the AMR of \citet{do10} with filled circles indicating 
$\rgc \leq 8$ kpc and diamonds indicating $\rgc > 8$ kpc. The current sample are plotted as open diamonds to 
highlight where they fall on the AMR. In order to give a sense of the uncertainties in the diagram, a `typical'
error bar representing $\pm0.2$ dex in [Fe/H] and $\pm1$ Gyr in age is plotted in every panel of Figure \ref{AMR}.
The remaining panels compare the GC AMR to various model predictions: the upper-right panel shows
the simple closed-box chemical evolution model for the Sagittarius dwarf spheroidal galaxy from \citet{si07}; 
the lower-left panel shows the chemical evolution models of the Magellanic Clouds from \citet{pa98}; and
the lower-right panel shows the mean trend from the GC formation model of \citet{mg10}. These comparisons are 
discussed in the following paragraphs.

The closed-box model portrayed in the upper-right panel of Figure \ref{AMR} was taken from the study of the star 
formation history of M\,54 and the central field of the Sagittarius (hereafter Sgr) dwarf spheroidal galaxy by
\citet{si07}. The star formation history was derived with the same stellar models used in this paper and photometry 
from the ACS Survey of Galactic GCs. It has already been demonstrated by \citet{ls00} that the GCs associated with Sgr 
follow the same trend as its field stars. \citet{fo10} reiterate that the AMR of Sgr is essentially identical to that of 
the putative Canis Major (CMa) dwarf galaxy and argue that $\sim 20$ GCs originated with one or the other. The upper-right panel 
of Figure \ref{AMR} indicates that the outer halo GCs follow an AMR consistent with dwarf galaxy chemical evolution. 
However, it appears that the outer halo GCs formed in a variety of environments within which chemical enrichment proceeded 
at different rates; one characteristic of Sgr or CMa is insufficient to explain the observed AMR.

The GC AMR is compared to the AMRs of the Small and Large Magellanic Cloud (SMC and LMC) 
chemical evolution models by \citet{pa98} in the lower-left panel of Figure \ref{AMR}. The LMC AMR is not markedly 
different from that of M\,54/Sgr, as pointed out by \citet{fo10}. The SMC AMR tracks the majority of GCs with ages less 
than 12 Gyr better than either the LMC or M\,54/Sgr.

The comparisons with Sgr and the MCs indicate that a collection of dwarf galaxies with a range of AMRs are likely to
have contributed to the Galactic GC population, consistent with the findings of \citet{fo10}.
\citet{mg10} devised a semi-analytical model of the formation of a massive host galaxy's GC population and trained their
model on the Galactic GC metallicity distribution. The model is built on top of cold dark matter galaxy formation 
simulations and assigns a globular cluster formation probability and chemical enrichment history to each sub-halo
that is accreted by the host galaxy. Among their stated goals is the development of a bimodal color distribution in 
the GC population without requiring two distinct formation mechanisms. While the model achieves this goal by design,
it also fits the age-metallicity distribution of the outer halo GCs remarkably well. The lower-right panel of Figure 
\ref{AMR} compares the GC AMR with the mean trend (squares) and standard deviation (indicated by error bars) of [Fe/H] for 
0.5 Gyr age bins of nearly 10,000 simulated GCs drawn from the \citet{mg10} model. While the outer halo GCs are well 
represented by the mean trend, the model is discrepant with the metal-rich GCs in the inner Galaxy, i.e., the bulge 
and thick disk GCs. Muratov \& Gnedin attribute this discrepancy to the lack of a metallicity gradient within the 
individual merging halos. As a substantial fraction of the Galactic GC population is old and metal-rich, it behooves 
a model of the GC population formation to describe the origins of these GCs.

\section{Discussion}

The AMRs constructed by \citet{mf09} and \citet[][further updated in this paper]{do10} show that Galactic GCs split into
two distinct branches at [Fe/H] $\ga -1.5$. \citet{mf09} interpreted this branching as representing two groups of GCs: 
an old group with uniformly old ages (a flat AMR) and a young group with a significant trend toward younger ages at 
higher metallicities. They concluded that these groups likely originated from two different phases of Galaxy formation: 
an initial, relatively brief collapse and a second, prolonged episode driven by accretion. \citet{mf09} found a small
number of metal-poor ([M/H] $\le -1.5$) GCs with somewhat younger ages than the bulk of the metal-poor GCs. \citet{do10}
and the present study find all of the metal-poor GCs to be uniformly old,\footnote{See Figure 9 of \citet{do10} for a 
direct comparison of the AMRs.} but this difference does not substantially influence the conclusion with respect to the 
proposed Galaxy formation scenario.

A number of influential studies have used the HB morphology-metallicity diagram to infer the ages of stellar 
populations for which metallicity information and good photometry to at least the level of the HB are available. 
For example, \citet{ca93} and \citet{ldz94} both used synthetic HB models to demonstrate that, under the assumption 
of constant mass loss, the Galactic GC HB morphology-metallicity diagram 
showed evidence for an age spread of several ($\sim4$) Gyr for a fixed He content and heavy element distribution. 

The use of the HB morphology-metallicity diagram  must, however, remain a crude age indicator at best for a variety of reasons.  
The lack of detailed abundance information, including the possible spread in He within a given GC, has important implications for 
HB morphology. The effect of a spread in the total He content on HB morphology \emph{among} different GCs has been explored 
by many authors \citep[see, e.g., ][]{ca93,ldz94} as well as a spread in He \emph{within} single GCs \citep[see, e.g., ][]{dc08}.
Another important factor in matching GC HB morphology to model predictions is statistical fluctuations. The ACS GC Treasury project
photometric database contains GCs with anywhere from $\sim20$ to several hundred HB stars, leading to uncertainty in the median
color of the HB of up to 30\% in the sparsest GC \citep[][Table 1]{do10}.

Addressing the question of which parameters most strongly influence HB morphology using full CMDs for large samples of GCs, 
\citet{do10} and \citet{gr10} reached the same conclusion: 
that age is the second parameter influencing HB morphology. The results presented in this paper lend further support to this 
conclusion. It is important to note that the term `second parameter' used here applies specifically to the variation of HB 
morphology \emph{among} different GCs, the context in which the term was originally used in the 1960's \citep[e.g., ][]{vdb67,sa67}. 
`Second parameter' is not intended to apply to the spread in HB stars within a single GC, for which the same term is often employed, 
nor does this paper seek to address that important issue. \citet{do10} and \citet{gr10} both found evidence for at least a third 
parameter as well, suggesting that age and metallicity alone are not sufficient to fully characterize HB morphology.

\begin{figure}
\plotone{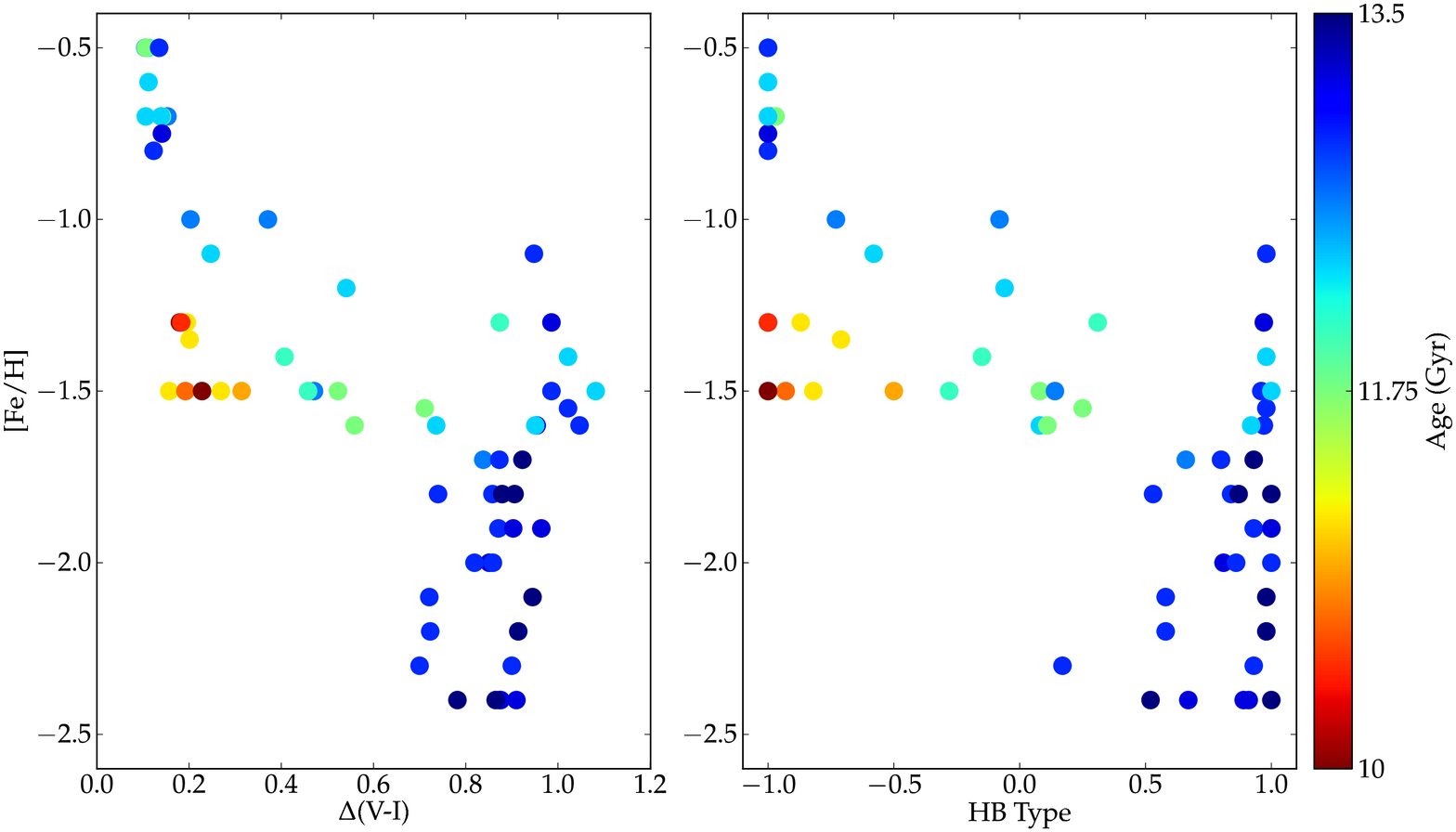}
\caption{A demonstration of the age effect in the HB morphology-metallicity diagram using ages derived from the
MSTO. The left panel shows the $\dvi$ metric as compiled by \citet{do10} and this study; the right panel shows (B$-$R)/(B+V+R)
as compiled by \citet{ma05}. The shade of each point corresponds to its age with the oldest GCs shown as blue
and the youngest shown as red.\label{HB}}
\end{figure}

Figure \ref{HB} shows the HB morphology-metallicity diagram of the full sample of 68 GCs. The left panel
considers HB morphology as the median color difference between the HB and RGB \citep{do10} and the right panel
instead uses the (B$-$R)/(B+V+R) metric of \citet{le89} as compiled by \citet{ma05}. \citet{do10} demonstrated
that these two HB morphology metrics are strongly correlated, with the main difference being that $\dvi$ does
not saturate as (B$-$R)/(B+V+R) does at $\pm1$. In this context, it is clear that age is the second parameter 
influencing HB morphology as concluded by \citet{do10} and \citet{gr10}.

While Figure \ref{HB} is not well sampled over the full range of metallicity, it can still be seen that the 
transition from a red HB to a blue one happens over $\la 0.5$ dex in [Fe/H] for a fixed age, or $\la 2$ Gyr at fixed
[Fe/H]. It is interesting to consider the ensemble in this manner, despite the fact that age uncertainties remain
at or near the 10\% level. It is a fortunate coincidence that we observe these GCs at a time when the HB 
morphology-metallicity diagram is rich with information. Turn the clock back $\sim2$ Gyr and the HB 
morphology-metallicity diagram would have been populated only on the red side---as, for example, in the 
SMC \citep{gl08}. At such a time, the study by \citet{sz78} would have revealed little variation in HB morphology 
among the halo GCs and, hence, little or no insight into the formation of the Galaxy. Only the truly peculiar GCs, 
such as NGC~2808 \citep{da10}, might have shown a sizable variation. On the other hand, turn the clock forward 
$\sim2$ Gyr and all but the most metal-rich Galactic GCs would have blue HBs.

\section{Summary}

HST/ACS photometry of 6 Galactic halo GCs, IC~4499, NGC~6426, NGC~7006, Palomar~15, Pyxis, and Ruprecht~106,
were presented. The resulting CMDs were used to derive ages via isochrone fitting. The age of Palomar~5 
was derived using the same set of models and the combined photometry of \citet{gr01} and \citet{st00}. The 
ages and metallicities of these 7 GCs were added to the AMR of \citet{do10}. The total number of homogeneously 
studied GCs in the sample is 68, excluding those 7 GCs imaged by the ACS Survey of Galactic GCs that are known 
to harbor multiple, distinct stellar populations. Divided at $\rgc = 8$ kpc, the inner Galaxy GCs exhibit a pattern 
of rapid chemical enrichment spanning two orders of magnitude in metallicity over a timescale of 1-2 Gyr. The outer 
Galaxy GCs exhibit a much slower chemical enrichment that is evocative of dwarf galaxy chemical evolution and is 
reasonably well matched by the semi-analytic GC formation models of \citet{mg10}.

\acknowledgments
Data presented in this paper were obtained from the Multimission Archive at the Space 
Telescope Science Institute (MAST). Support for this work (proposal GO-11586) was provided by NASA 
through a grant from the Space Telescope Science Institute, which is operated by the Association of 
Universities for Research in Astronomy, Inc., under NASA contract NAS5-26555. This research has made 
use of NASA's Astrophysics Data System Bibliographic Service as well as the SIMBAD database, operated 
at CDS, Strasbourg, France.

\end{document}